\documentclass[11pt]{article}
\usepackage{graphicx}
\begin{document}

\title{The Connection of Polaritons and Vacuum Rabi Splitting}
\author{David Snoke\\
Department of Physics and Astronomy, University of Pittsburgh\\
3941 O'Hara St., Pittsburgh, PA 15260}
\date{}

\maketitle

\abstract{Polaritons, in particular microcavity exciton-polaritons, have attracted much attention in recent years, as the phenomena of Bose-Einstein condensation and superfluidity have been observed for these quasiparticles. While the basic physics of these systems is well understood, there has been confusion over the connection of these systems to other phenomena, namely the Jaynes-Cummings Hamiltonian, Rabi flopping, and the vacuum Rabi splitting of atoms in a cavity. This paper reviews the basic theory of polaritons and shows these connections explicitly.}

\section{Introduction}

Numerous recent works \cite{deng,caru,keeling,snoke-chap} have reviewed the fascinating effects seen in experiments on Bose-Einstein condensation and superfluidity of microcavity exciton-polaritons. The field has progressed from simply observing evidence of Bose-Einstein condensation \cite{dev,balili} to various effects such as Josephson junctions \cite{jo1,jo2}, phase locking of two condensates \cite{phase}, superfluid motion past a barrier \cite{superfluid}, and quantized circulation in a ring \cite{pnas}. The lifetime of the polaritons in the cavities has steadily progressed, from about a picosecond in early experiments up to about 200 ps in present systems \cite{turnaround}.

Amidst all this activity, some basic confusions remain about the nomenclature for the system and the connection to other systems with light-matter coupling. For example, does the ``Rabi splitting'' of exciton-polariton systems have anything to do with Rabi oscillations or vacuum Rabi splitting? We will see here that indeed these are variations of the same phenomenon. Similarly, is the term ``strong coupling'' used in the polariton field in the same way as in other fields of optics? Again, we will see that this is the case.

The basic concept of a polariton is simple. One starts with any oscillating dipole which can couple to the electromagnetic field. This can be, for example, an optical phonon in a solid, an exciton in a solid, or a two-level quantum oscillator consisting of two states in an atom. The coupling of the dipole oscillation to the electromagnetic field allows both radiation from the dipole or absorption of radiation by the dipole.

\section{Phonon-polaritons}

The polariton effect is most easily introduced by examining the case of phonon-polaritons, when an optical phonon in a solid couples to the electromagnetic field. In this section we review the standard theory for phonon-polaritons.

We start with the interaction energy of a dipole in an external electric field, $H_{\rm dipole} = -q\vec{x}\cdot\vec{E}$.  For a polarization field $\vec{P} = (N/V)q\vec{x}$, the total interaction energy is then
\begin{equation}
H_{\rm int} = - \int d^3 r \  \vec{P}\cdot \vec{E} .
\label{polhint}
\end{equation}
The polarization, which is proportional to the local displacement, and the electric field can be written in terms of the phonon and photon operators. Dropping the vector notation for simplicity, the electric field is \cite{snokebook-phot}
\begin{equation}
E(r) = -i\sum_k \sqrt{\frac{\hbar\omega_k}{2\epsilon_\infty V }}\left(a_k^{ }e^{ik\cdot r}-a_{k}^{\dagger}e^{-ik\cdot r}\right),
\label{photE}
\end{equation}
where $\epsilon_{\infty}$ is the dielectric constant of the medium not counting the contribution of the optical phonons, and the polarization field is \cite{snokebook-phon}
\begin{equation}
P(r) = \frac{N}{V}q x(r) = \frac{qN}{V} \sqrt{\frac{\hbar}{2 m_r N \omega_0}} \sum_k \left(c_k^{ }e^{ik\cdot r} + c_{k}^{\dagger}e^{-ik\cdot r}\right) ,
\end{equation}
where we have used creation and destruction operators $c_k^{\dagger}$ and  $c_k$ for the phonons, and we assume an optical phonon  with constant frequency $\omega_0$. This may be written more simply as
\begin{equation}
P(r) = \sqrt{\frac{\hbar \epsilon_\infty\Omega^2}{2V\omega_0}}\sum_k \left(c_k^{ }e^{ik\cdot r} + c_{k}^{\dagger}e^{-ik\cdot r}\right) ,
\end{equation}
where $\Omega = \sqrt{q^2N/\epsilon_\infty m_r V}$. 
Substituting these formulas for $E$ and $P$ into $H_{\rm int}$, we obtain
\begin{equation}
H_{\rm int} = i\frac{\hbar\Omega}{2} \ \sqrt{\frac{\omega_k}{\omega_0}}\  \frac{1}{V}\sum_{k,k'}\int d^3r  \left(a_k^{ }e^{ik\cdot r}-a_{k}^{\dagger}e^{-ik\cdot r}\right) \left(c_{k'}^{ }e^{ik'\cdot r} + c_{k'}^{\dagger}e^{-ik'\cdot r}\right) .
\end{equation}
The integral of the exponential factors over $\vec{r}$ gives us a $\delta_{k,k'}$ which eliminates one momentum sum, so that we have the total Hamiltonian
\begin{equation}
H = \sum_k \left(\hbar\omega_{k }  a_k^{\dagger}a^{ }_k + \hbar\omega_{0 } c_k^{\dagger}c^{ }_k + \frac{i}{2}\hbar\Omega \ \sqrt{\frac{\omega_k}{\omega_0}} \left(a^{ }_k c^{ }_{-k}  + a^{ }_k c_{k}^{\dagger} - a_k^{\dagger} c^{ }_{k} - a_k^{\dagger} c_{-k}^{\dagger}\right)\right) .
\end{equation}

This can be simplified by defining new operators which are linear superpositions of the creation and destruction operators which appear in the Hamiltonian. We define the new bosonic destruction operator $\xi_{k} = \alpha_{k}a_k^{ } + \beta_{k}c_k^{ } + \gamma_k a_{-k}^{\dagger} + \delta_k c_{-k}^{\dagger}$ and its Hermitian conjugate for the creation operator, with the coefficients $\alpha_{k}\ldots \delta_{k}$ chosen such that
$$
H = \sum_k \hbar\omega \ \xi_{k}^{\dagger}\xi^{ }_{k} ,
$$ 
which implies 
\begin{eqnarray}
[\xi_k, H ]&=&\hbar\omega_{ } \xi_k \nonumber\\
&=& \alpha_{k}[a_k,H] + \beta_{k}[c_k,H] + \gamma_k [a_{-k}^{\dagger},H] + \delta_k [c_{-k}^{\dagger},H] \nonumber\\
&=& \alpha_{k}\hbar\left(\omega_k a_k - \frac{i}{2}\hbar\Omega \sqrt{\frac{\omega_k}{\omega_0}}c_k - \frac{i}{2}\hbar\Omega \sqrt{\frac{\omega_k}{\omega_0}}c_{-k}^\dagger\right)\nonumber\\
&& + \beta_{k}\hbar\left(\omega_0 c_k + \frac{i}{2}\hbar\Omega \sqrt{\frac{\omega_k}{\omega_0}}a_k - \frac{i}{2}\hbar\Omega \sqrt{\frac{\omega_k}{\omega_0}}a_{-k}^\dagger\right)\nonumber\\
&& + \gamma_{k}\hbar\left(-\omega_k a^{\dagger}_{-k} - \frac{i}{2}\hbar\Omega \sqrt{\frac{\omega_k}{\omega_0}}c_k - \frac{i}{2}\hbar\Omega \sqrt{\frac{\omega_k}{\omega_0}}c_{-k}^\dagger\right)\nonumber\\
&& + \delta_{k}\hbar\left(-\omega_0 c^{\dagger}_{-k} - \frac{i}{2}\hbar\Omega \sqrt{\frac{\omega_k}{\omega_0}}a_k - \frac{i}{2}\hbar\Omega \sqrt{\frac{\omega_k}{\omega_0}}a_{-k}^\dagger\right).
\label{poldiag}
\end{eqnarray}

The condition (\ref{poldiag}) is equivalent to the matrix equation
\begin{eqnarray}
\left(
\begin{array}{cccc}
\omega_k & \frac{i}{2}\Omega\sqrt{{\omega_k}/{\omega_0}} & 0 & - \frac{i}{2}\Omega\sqrt{{\omega_k}/{\omega_0}}\\
- \frac{i}{2}\Omega\sqrt{{\omega_k}/{\omega_0}} & \omega_0 & - \frac{i}{2}\Omega\sqrt{{\omega_k}/{\omega_0}} & 0 \\
0 & - \frac{i}{2}\Omega\sqrt{{\omega_k}/{\omega_0}} & -\omega_k &  \frac{i}{2}\Omega\sqrt{{\omega_k}/{\omega_0}}\\
- \frac{i}{2}\Omega\sqrt{{\omega_k}/{\omega_0}} & 0 & - \frac{i}{2}\Omega\sqrt{{\omega_k}/{\omega_0}} & -\omega_0
\end{array}
\right)
\left(
\begin{array}{c}
\alpha_k\\
\beta_k\\
\gamma_k\\
\delta_k
\end{array}
\right)
=
\omega \left(
\begin{array}{c}
\alpha_k\\
\beta_k\\
\gamma_k\\
\delta_k
\end{array}
\right). \nonumber\\
\end{eqnarray}

Following the standard diagonalization procedure of setting the determinant to zero, we have
\begin{equation}
\omega^4 -\omega^2(\omega_0^2  + \omega_k^2) +\omega_k^2\omega_0^2 -\Omega^2\omega_k^2  = 0.
\end{equation}
This is equivalent to the standard phonon-polariton equation \cite{snokebook-pol}
\begin{equation}
\omega^2 = {c^2k^2 \over \epsilon({\infty})} \left( {\omega_T^2-
\omega^2 \over \omega_L^2-
\omega^2}\right),
\label{phpolariton}
\end{equation}
 with the bare photon frequency $\omega_k = ck/\sqrt{\epsilon_\infty}$ and $\omega_T^2 = \omega_L^2 - \Omega^2$, using 
\begin{equation}
{q^2 N \over mV} = \epsilon({\infty})(\omega_L^2- \omega_{T}^2 ).
\label{lsta}
\end{equation} 
and our definition of $\Omega$ above. There are two positive-frequency solutions.

When $\omega_k = \omega_0$, it is easy to show that in the limit $\omega_0 \gg \Omega$, the energies of the eigenstates are $\hbar\omega = \hbar\omega_0 \pm {\hbar\Omega}$, and the corresponding eigenstates are $\xi_k = (a_k \mp ic_k)/\sqrt{2}$.  Away from the crossover region, when either $\omega_k \ll \omega_0$ or $\omega_k \gg \omega_0$, the eigenmodes correspond to nearly pure photon $a_k$ and phonon $c_k$ operators. 

\section{Exciton-polaritons: ensemble of two-level oscillators\\ (Frenkel limit)}

We adopt the same picture as the phonon picture, but now imagine an ensemble of two-level electron oscillators at discrete locations $i$. The polarization of each oscillator is \cite{snokebook-coh}
\begin{eqnarray}
\vec{P}_i =  \frac{qN}{V}\vec{x}_i = \frac{qN}{V} 
\left( \langle c | \vec{x} | v\rangle  b^\dagger_{iv} b_{ic}^{ } + \langle c | \vec{x} | v\rangle^* b^\dagger_{ic} b_{iv}^{ } \right),
\end{eqnarray}
where we use $b_{in}^\dagger$ and $b_{in}^{ }$ for the fermionic creation and destruction operators of the electrons, and $n = c, v$ represent the conduction and valence bands, respectively; since in the Frenkel picture, each oscillator has no spatial overlap with the others, we could equally well call these $e$ and $g$ for the excited and ground states of each oscillator, as in the case of an ensemble of atoms.

We define the exciton operator in the Frenkel limit as 
\begin{equation}
C^\dagger_k = \frac{1}{\sqrt{N}}\sum_i e^{ik\cdot r_i} b^\dagger_{ic} b_{iv}^{ }.
\end{equation}
The inverse Fourier transform is
\begin{equation}
b^\dagger_{ic} b_{iv} = \frac{1}{\sqrt{N}}\sum_k e^{-ik\cdot r_i} C^\dagger_k,
\end{equation}
so we have
\begin{eqnarray}
\vec{P}_i = -i\frac{qN}{V}\frac{1}{m\omega_0 \sqrt{N}}\sum_k \left( p_{cv} C_k e^{ik\cdot r_i} - p_{cv}^* C^\dagger_k e^{-ik\cdot r_i} \right),
\end{eqnarray}
where we write $p_{cv} = \langle c | \vec{p} | v\rangle = im\omega_0 \langle c | \vec{x} | v\rangle$ \cite{snokebook-osc}. 
We then have
\begin{eqnarray}
H_{\rm int} &=&  - \int d^3 r \  \vec{P}\cdot \vec{E} \nonumber \\
&=& \sqrt{\frac{\hbar\omega_k}{2\epsilon V}}  \frac{qN}{V} \frac{1}{m\omega_0\sqrt{N}}\sum_{k,k'}\int d^3r  \left(a_ke^{ik\cdot r}-a_{k}^{\dagger}e^{-ik\cdot r}\right)  \left(p_{cv}^* C_{k'}^{\dagger}e^{-ik'\cdot r}-p_{cv} C_{k'}e^{ik'\cdot r} \right) \nonumber\\
&=& \frac{\hbar}{2}\sqrt{\frac{ q^2 N}{\epsilon m V}}   \sqrt{\frac{2}{m\hbar\omega_0}}\sqrt{\frac{\omega_k}{\omega_0}}  \sum_{k} \left(-p_{cv} a_kC_{-k}+  p_{cv}^* a_kC_{k}^{\dagger}+p_{cv}a^\dagger_kC_k - p_{cv}^* a^\dagger_kC_{-k}^{\dagger}\right). \nonumber\\ \label{frenkint}
\end{eqnarray}

As for the phonon-polaritons, we write $\xi_{k} = \alpha_{k}a_k^{ } + \beta_{k}C_k^{ } + \gamma_k a_{-k}^{\dagger} + \delta_k C_{-k}^{\dagger}$. Then
\begin{eqnarray}
H &=& \sum_k \hbar\omega \xi^\dagger_k \xi_k^{ }  = \sum_k (\hbar \omega_k a^\dagger_k a_k^{ } + \hbar\omega_0 C^\dagger_kC_k^{ }) + H_{\rm int}
\end{eqnarray}
and
\begin{eqnarray}
[\xi_k, H ]&=&\hbar\omega_{ } \xi_k^{ } \nonumber\\
&=& \left( \alpha_{k}[a_k,H] + \beta_{k}p_{cv}[C_k,H] + \gamma_k [a_{-k}^{\dagger},H] + \delta_k p_{cv}^*[C_{-k}^{\dagger},H]\right) \nonumber\\
&=& \hbar \alpha_{k}\left(  \omega_k a_k +\frac{\Omega}{2}\sqrt{\frac{\omega_k}{\omega_0}}C_k-\frac{\Omega^*}{2}\sqrt{\frac{\omega_k}{\omega_0}}C^\dagger_k\right)\nonumber\\
&&+ \hbar\beta_k \left(\omega_0C_k  +\frac{\Omega^*}{2}\sqrt{\frac{\omega_k}{\omega_0}}a_k-\frac{\Omega^*}{2}\sqrt{\frac{\omega_k}{\omega_0}}a^\dagger_{-k}\right) \nonumber\\
&&+\hbar\gamma_k\left( - \omega_k a^\dagger_{-k} +\frac{\Omega}{2}\sqrt{\frac{\omega_k}{\omega_0}}C_{k}-\frac{\Omega^*}{2}\sqrt{\frac{\omega_k}{\omega_0}}C^\dagger_{-k}\right) \nonumber\\
&&+\hbar\delta_k \left(-\omega_0C^\dagger_{-k}  +\frac{\Omega}{2}\sqrt{\frac{\omega_k}{\omega_0}}a_{k}-\frac{\Omega}{2}\sqrt{\frac{\omega_k}{\omega_0}}a^\dagger_{-k}\right),
\end{eqnarray}
with 
\begin{equation}
\Omega = \sqrt{\frac{q^2 N}{\epsilon m V}} F
\label{frenkelomega}
\end{equation}
and $F = p_{cv}\sqrt{2/m\hbar\omega_0}$; $|F|^2$ is the oscillator strength. (We have used the approximation that the $C_k$ operators are purely bosonic.)

This is equivalent to the matrix equation
\begin{equation}
\left(
\begin{array}{cccc}
\omega_k & \frac{1}{2}\Omega^* \sqrt{{\omega_k}/{\omega_0}}& 0 &  \frac{1}{2}\Omega\sqrt{{\omega_k}/{\omega_0}}\\
 \frac{1}{2}\Omega\sqrt{{\omega_k}/{\omega_0}} & \omega_0 &  \frac{1}{2}\Omega \sqrt{{\omega_k}/{\omega_0}}& 0 \\
0 &  -\frac{1}{2}\Omega^*\sqrt{{\omega_k}/{\omega_0}} & -\omega_k & - \frac{1}{2}\Omega\sqrt{{\omega_k}/{\omega_0}}\\
 -\frac{1}{2}\Omega^*\sqrt{{\omega_k}/{\omega_0}} & 0 &  -\frac{1}{2}\Omega^* \sqrt{{\omega_k}/{\omega_0}}& -\omega_0
\end{array}
\right)
\left(
\begin{array}{c}
\alpha_k\\
\beta_k\\
\gamma_k\\
\delta_k
\end{array}
\right)
=
\omega \left(
\begin{array}{c}
\alpha_k\\
\beta_k\\
\gamma_k\\
\delta_k
\end{array}
\right).\\
\end{equation}
The determinant is
\begin{equation}
\omega^4 -\omega^2(\omega_0^2  + \omega_k^2) +\omega_k^2\omega_0^2 -|\Omega|^2\omega_k^2  = 0.
\label{frenkdiag}
\end{equation}
This is equivalent to the standard polariton equation
\begin{equation}
\omega^2 = \frac{c^2k^2}{\epsilon_\infty}\left( {\omega_T^2-
\omega^2 \over \omega_0^2-
\omega^2}\right)
\end{equation}
with $\omega_T^2 = \omega_0^2 - \Omega^2$ and $\omega_k = ck/\sqrt{\epsilon_{\infty}}$.

%$$
%\omega^4 -2\omega^2\omega_0^2   +\omega_0^2(\omega_0^2 -|\Omega|^2)  = 0.
%$$
%$$
%\omega^2 = \omega_0^2 \pm\frac{1}{2} \sqrt{4\omega_0^4 - 4(\omega_0^4-\omega_0^2\Omega^2)}  = \omega_0^2 \pm \omega_0\Omega
%$$
%$$
%\omega = \sqrt{\omega_0^2\pm \omega_0\Omega} = \omega_0\sqrt{1\pm \Omega/\omega_0} \simeq \omega_0 \pm \Omega/2
%$$
%$$
%\omega_T = \sqrt{\omega_0^2 - \Omega^2} = \omega_0\sqrt{1-\Omega^2/\omega_0} \simeq \omega_0 (1-\Omega^2/2\omega_0)
%$$

%We can write $\Omega = \sqrt{q^2 /\epsilon m a^3}$, where $a$ is the unit cell size (which is the size of a Frenkel exciton). 

{\bf Cavity Polaritons}. In a cavity we use $\omega_k = (c/n)\sqrt{k_{\|}^2 + k_{\perp}^2}) \simeq (c/n)(1+k_{\|}^2/2k_{\perp})$, with $k_{\perp} = \pi/L$. For the resonant case we set $\omega_k = (c/n)\pi/L = \omega_0$. Then the determinant equation is
\begin{equation}
\omega^4 -2\omega^2\omega_0^2   +\omega_0^2(\omega_0^2 -|\Omega|^2)  = 0,
\end{equation}
which has the solutions
\begin{equation}
\omega^2 = \omega_0^2 \pm \omega_0 \Omega,
\end{equation}
or 
\begin{equation}
\omega = \omega_0 \sqrt{1 \pm \Omega/\omega_0} \simeq \omega_0 \pm \frac{\Omega}{2},
\end{equation}
where the final approximation is valid when $\Omega \ll \omega_0$. These correspond to the lower polaritons and upper polaritons so familiar in microcavity polariton optics. There will also be ``dark,'' uncoupled states at $\omega_0$ which lie in the middle of the upper-lower polariton splitting. This uncoupled $\omega_0$ was $\omega_L$ in the case of the phonon-polaritons. 

%%%%%%%%%%consistency of definition of $\Omega$?

%$$
%\omega^2 ( \omega_L^2- \omega^2) - \omega_k^2\left( \omega_T^2- \omega^2 \right) = 0,
%$$
%$$
%\omega^4 -\omega^2  \omega_L^2  + \omega_k^2 \omega_T^2- \omega_k^2\omega^2  = 0
%$$
%$$
%\omega^4 -\omega^2  (\omega_L^2+\omega_k^2)  + \omega_k^2 \omega_T^2  = 0
%$$
%$$
%\omega^4 -\omega^2  (\omega_L^2+\omega_k^2)  + \omega_k^2( \omega_0^2-\Omega^2)  = 0
%$$

\section{Exciton-polaritons: Wannier picture}
\label{sect.wannier}

The Wannier limit of excitons consists of the case in which the electron and hole (empty electron state) are no longer confined to the same oscillator, but instead, due to coupling between the oscillators, the electron and hole can migrate to different oscillators. In this case there will be a Coulomb attraction between the free electron and the hole, which effectively acts as a particle with positive charge. This leads to bound states of the free electron and hole that are exactly the same as the Rydberg bound states of a hydrogen atom, but with the energy scaled by the dielectric constant of the medium. 

For this calculation we use the interaction Hamiltonian in $k$-space instead of in real space:
\begin{eqnarray}
 H_{\rm int} &=&  -\frac{q}{m}  \sum_{{k,k'}}  
\sqrt{\frac{\hbar}{2\epsilon V\omega_k}}  \biggl[ \langle c | \vec{p} | v
\rangle \left( a^{ }_{{k}}  b^{\dagger}_{c,{k}'+{k}}b^{ }_{v{k}'} + 
a^{\dagger}_{{k}}b^{\dagger}_{c,{k}'-{k}}b^{ }_{v,{k}'}\right) + \nonumber\\
&&+ \langle c | \vec{p} | v
\rangle^*\left(a^{ }_{{k}}  b^{\dagger}_{v,{k}'+{k}}b^{ }_{c{k}'} + 
a^{\dagger}_{{k}}b^{\dagger}_{v,{k}'-{k}}b^{ }_{c,{k}'} \right)\biggr].
\label{dipoleint}
\end{eqnarray}
The exciton creation operator in the Wannier case is \cite{HH}
\begin{eqnarray}
C^{\dagger}_k &=&  \sum_{k'} \phi(k/2-k') b^\dagger_{c,k-k'}b^{ }_{v,-k'}.
\end{eqnarray}
% the Gaussian 1s wave function is  
%$$
%\phi(r) = \frac{e^{-r/a}}{\sqrt{\pi a^3}}.
%$$
%We thus have
%\begin{equation}
%C^{\dagger}_k = \frac{1}{\sqrt{N}}\sum_{i,j} b^\dagger_{c,i}b^{ }_{v,j} e^{ik\cdot r_i} \left(\sqrt{\frac{V}{N}}\frac{e^{-|r_i-r_j|/a}}{\sqrt{\pi %a^3}}\right).
%\end{equation}
The Fourier transform of the 1s wave function is
\begin{equation}
\phi(k) =\frac{1}{\sqrt{V}} \frac{8\sqrt{\pi a^3}}{(1+a^2k^2)^2},
\end{equation}
where $a$ is the exciton Bohr radius.

It is not easy to invert this to write the Hamiltonian in terms of the exciton operators. Therefore instead of an exact diagonalization, we write a matrix on the states $|{\rm ex} \rangle = C^\dagger_k|0\rangle$ and $|{\rm phot} \rangle = a^\dagger_k|0\rangle$. The off-diagonal term is
\begin{eqnarray}
\langle {\rm ex}|H|{\rm phot} \rangle &=& - \langle 0 | \sum_{k'''} \phi(k/2-k''') b^{\dagger}_{v,-k'''} b_{c,k-k'''} \frac{q}{m}  \sum_{{k'',k'}}  
\sqrt{\frac{\hbar}{2\epsilon V\omega_{k''}}} \\
&&\times \biggl[ \langle c | \vec{p} | v
\rangle \left( a^{ }_{{k}''}  b^{\dagger}_{c,{k}'+{k}''}b^{ }_{v{k}'} + 
a^{\dagger}_{{k}''}b^{\dagger}_{c,{k}'-{k}''}b^{ }_{v,{k}'}\right)  \\
&& + \langle c | \vec{p} | v
\rangle^*\left(a^{ }_{{k}''}  b^{\dagger}_{v,{k}'+{k}''}b^{ }_{c{k}'} + 
a^{\dagger}_{{k}''}b^{\dagger}_{v,{k}'-{k}''}b^{ }_{c,{k}'} \right)\biggr] a_k^\dagger|0\rangle 
\nonumber\\
&=& - \langle 0 | \sum_{k'''} \phi(k/2-k''') b^{\dagger}_{v,-k'''} b_{c,k-k'''} \\
&&\times \frac{q}{m}  \sum_{{k'}}  
\sqrt{\frac{\hbar}{2\epsilon V\omega_{k}}}  \biggl[ \langle c | \vec{p} | v
\rangle  b^{\dagger}_{c,{k}'+{k}}b^{ }_{v{k}'}  + \langle c | \vec{p} | v
\rangle^*  b^{\dagger}_{v,{k}'+{k}}b^{ }_{c{k}'}  \biggr] |0\rangle  \nonumber\\
&=& - \sum_{{k'}} \phi(k/2+k')  \frac{q}{m}    
\sqrt{\frac{\hbar}{2\epsilon V\omega_{k}}}   \langle c | \vec{p} | v
\rangle   .
\end{eqnarray}
The sum over $k'$ is
\begin{eqnarray}
&& \frac{V}{(2\pi)^3}\int 2\pi k'^2 dk' d(\cos\theta)\frac{1}{\sqrt{V}} \frac{8\sqrt{\pi a^3}}{(1+a^2|\vec{k}/2+\vec{k'}|^2)^2} \nonumber\\
&=& \frac{\sqrt{V}}{(2\pi)^3}\int 2\pi k'^2 dk' d(\cos\theta)\frac{8\sqrt{\pi a^3}}{(1+a^2(\frac{1}{4}k^2+{k'}^2+kk'\cos\theta))^2} \nonumber\\
&=&  \frac{\sqrt{V}}{(2\pi)^3}\int_0^\infty 2\pi k'^2 dk' \frac{32\sqrt{\pi a^3}}{2+\frac{1}{8}a^4(k^2-4k'^2)^2+a^2(k^2+4k'^2)}\nonumber\\
&=& \frac{\sqrt{V}}{\sqrt{\pi a^3}}.
\end{eqnarray}
Therefore
\begin{eqnarray}
\langle {\rm ex}|H|{\rm phot} \rangle &=& \frac{\hbar}{2} \frac{1}{\sqrt{\pi a^3}}      
\sqrt{\frac{ q^2}{\epsilon m }} \sqrt{\frac{\omega_0}{\omega_{k}} } \langle c | \vec{p} | v \rangle \sqrt{\frac{2}{m\hbar\omega_0}} \nonumber\\
&=& \frac{\hbar}{2}\Omega  \sqrt{\frac{\omega_0}{\omega_{k}} } ,
\end{eqnarray}
where 
\begin{equation}
\Omega = \sqrt{\frac{ q^2}{\epsilon m \pi a^3}}F
\label{wannier}
\end{equation}
and $F$ is defined as above for the oscillator strength. 

We thus have the matrix
\begin{eqnarray}
\left(
\begin{array}{cc}
\langle {\rm ex}|H|{\rm ex}\rangle & \langle {\rm ex}|H|{\rm phot}\rangle\\
\langle {\rm phot}|H|{\rm ex}\rangle & \langle {\rm phot}|H|{\rm phot}\rangle
\end{array}
\right) 
=
\hbar \left(
\begin{array}{cc}
\omega_0 & \frac{1}{2}\Omega \sqrt{\omega_0/\omega_k}\\
 \frac{1}{2}\Omega^* \sqrt{\omega_0/\omega_k} & \omega_k
\end{array}
\right)
\end{eqnarray}
which has the determinant equation
%$$
%(\omega_0-\omega)(\omega_k-\omega) -\frac{1}{4}|\Omega|^2\omega_0/\omega_k = 0 
%$$
%or
\begin{eqnarray}
\omega_0\omega_k -\omega(\omega_k + \omega_0) + \omega^2-\frac{1}{4}|\Omega|^2\omega_0/\omega_k = 0.
\label{wannierapprox}
\end{eqnarray}
This has the same behavior as the previous determinant equation (\ref{frenkdiag}) for the exact diagonalization, except near $\omega_k = 0$ where it breaks down. See Figures 1 and 2. 

Note that the Rabi frequency (\ref{wannier}) depends on the exciton Bohr radius $a$ through the volume $\pi a^3$.  The Frenkel limit can be viewed as the Wannier picture of excitons in the case when the exciton Bohr radius becomes equal to the unit cell size of the underlying crystal. This, in the Frenkel limit, the volume $\pi a^3$ becomes the unit cell size $a_L^3 = V/N$, in which case the Rabi frequency (\ref{wannier}) of the Wannier limit becomes exactly the same as (\ref{frenkelomega}) in the Frenkel limit.

\begin{figure}[h]
\begin{center}
\includegraphics[width=0.55\textwidth]{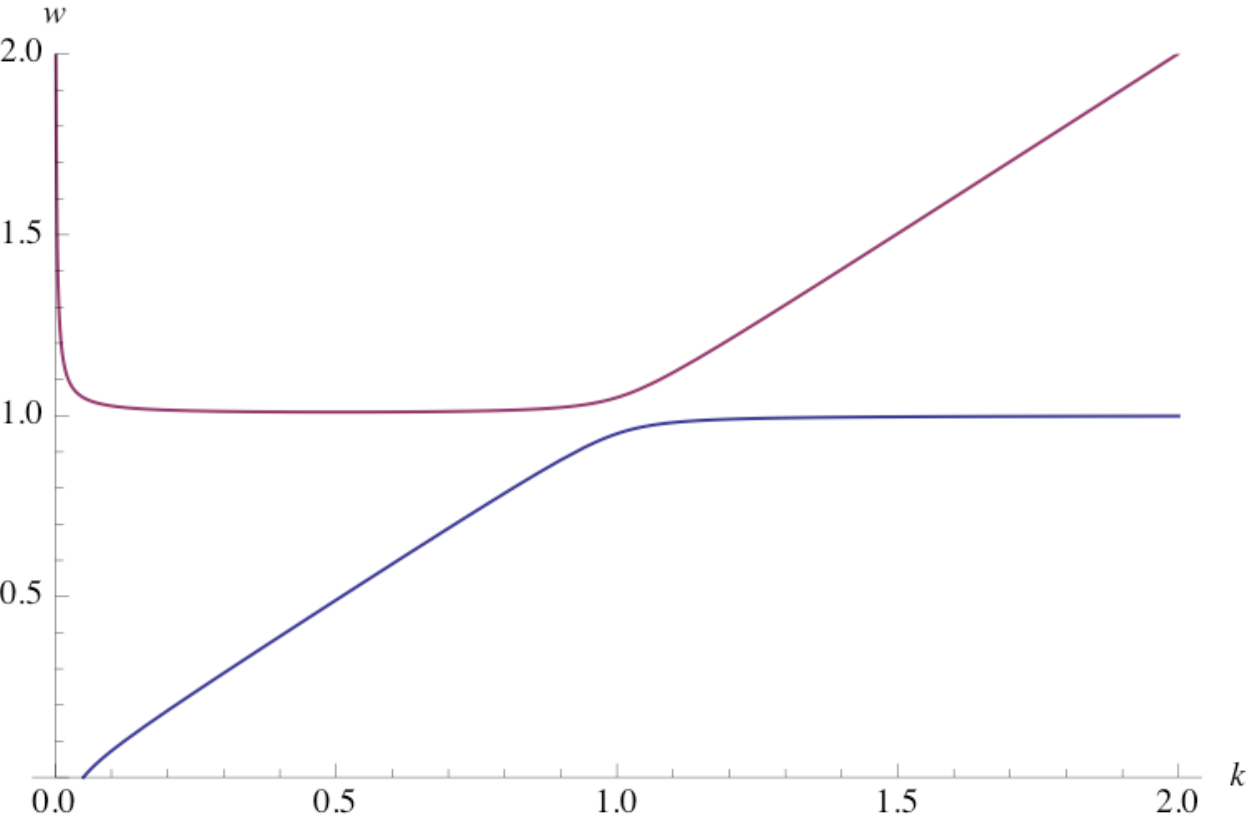}
\caption{Plot of the solutions of Eq.~(\protect\ref{wannierapprox}) for $\omega_0 = 1, \Omega = 0.1$.}
\end{center}
\end{figure}

\begin{figure}[h] 
\begin{center}
\includegraphics[width=0.55\textwidth]{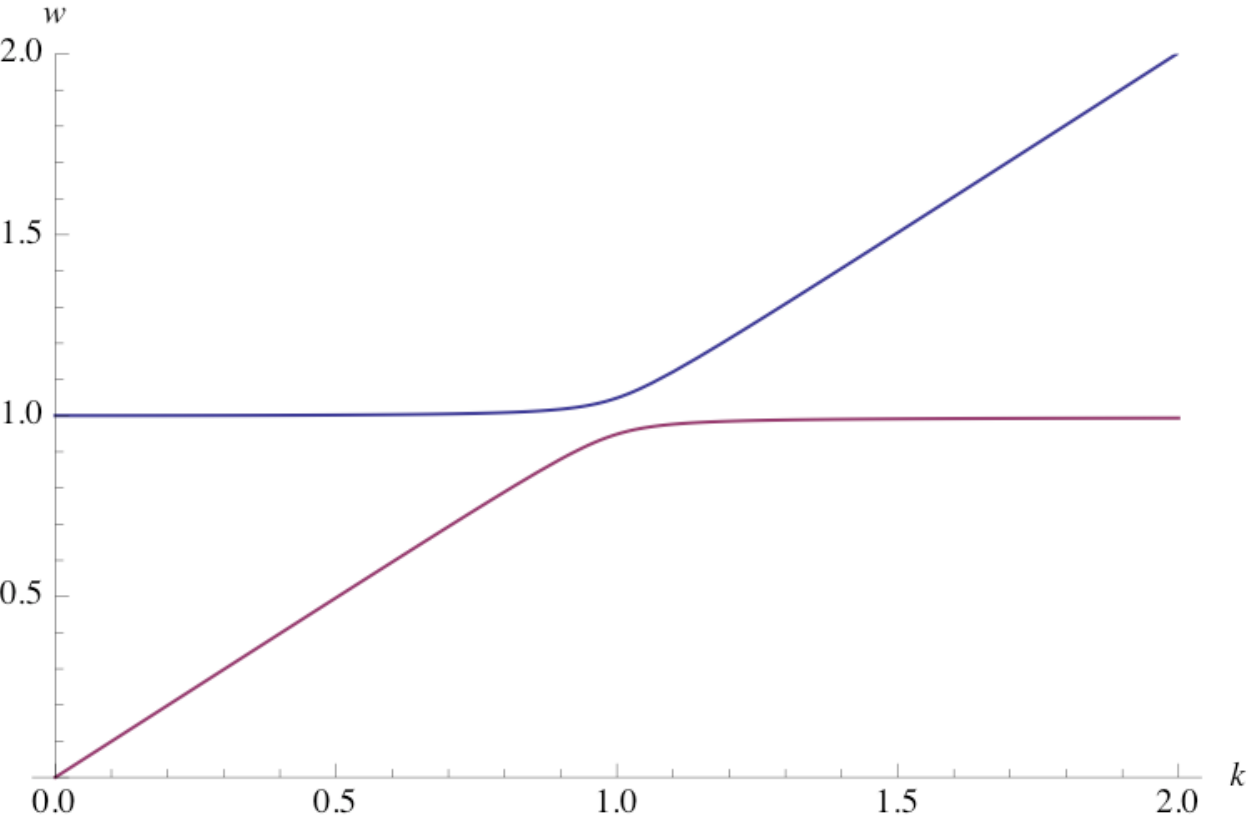}
\caption{Plot of the solutions of Eq.~(\protect\ref{frenkdiag}) for $\omega_0 = 1, \Omega = 0.1$.}
\end{center}
\end{figure}

\section{Comparison to atomic vacuum Rabi splitting}

As mentioned above, the Frenkel exciton limit is no different from the case of an ensemble of isolated atoms.  We can therefore compare the case of an ensemble of atoms in vacuum to the above results with just a small change in notation. 

The standard Hamiltonian for vacuum Rabi splitting with a two-level atom is \cite{agerwal}
\begin{equation}
H = \sum_i \hbar\omega_0 b^\dagger_{ic} b_{iv}^{ } + \hbar\omega_k a^\dagger_k a_k^{ } +\sum_i  \hbar g (a_k^{ } b^{\dagger}_{ic} b_{iv}^{ } + a^\dagger_kb^{\dagger}_{iv}b_{ic}^{ }),
\label{ager}
\end{equation}
where
\begin{eqnarray}
 g = \sqrt{\frac{\omega_0 d^2}{2\hbar\epsilon V}} = \sqrt{\frac{\omega_0 q^2\langle x\rangle^2}{2\hbar \epsilon V}} = \frac{1}{2}\sqrt{\frac{q^2 }{\epsilon m V}\left(\frac{2\langle p\rangle^2}{m\hbar\omega_0}\right)},
\end{eqnarray}
and we have again used $\langle p\rangle = im\omega \langle x\rangle$. The Hamiltonian (\ref{ager}) is equivalent to (\ref{frenkint}) with the rescaling of the photon operator
\begin{equation}
a_{k}^{ } \rightarrow ie^{ik\cdot r} a_k^{ },
\end{equation}
assuming $p_{cv}$ is real,
and dropping the two terms $a_k^{ } b^{ }_{ic} b_{iv}^{\dagger}$ and $a^\dagger_kb^{}_{iv}b_{ic}^{\dagger}$, which is comparable to the approximation made in Section~\ref{sect.wannier}, which is valid when $\omega_0 \gg \Omega$.

Agerwal \cite{agerwal} found
that the splitting is
\begin{equation}
\Omega = 2g\sqrt{N},
\label{ager2}
\end{equation}
 which is exactly the same as found for our definition (\ref{frenkelomega}) for the Rabi splitting of a two-level Frenkel exciton system. It really is the same system, namely an ensemble of independent two-level oscillators coupled only by the electromagnetic field, just solved in a different way. 

It may at first seem as though the result (\ref{wannier}) is not the same as the coupling (\ref{ager2}), since the latter is proportional to $\sqrt{N}$, where $N$ is the number of atoms, while (\ref{wannier}) is not proportional to the number of excitons. It is important to keep in mind, however, that the number of oscillators in an excitonic system is not given by the number of excitons, but by the number of atoms in the underlying medium. In the Frenkel case, the number of oscillators is exactly the number of atoms, while in the Wannier case, the effective number of oscillators is reduced by the ratio $(a_L/a)^3$, since each Wannier exciton is spread over many lattice sites.

\section{Final Remarks}

We have seen that the description of the interaction of electromagnetic field with all three types of electronic oscillation, namely Frenkel excitons, Wannier excitons, and exceptions of atomic states, can be described by the same formalism. The coupling term in each case can be viewed as the natural unit of frequency for the susceptibility of an ensemble of classical oscillators \cite{snokebook-classosc}
\begin{eqnarray}
 g \sim \sqrt{ \frac{q^2}{\epsilon m V_{\rm osc}}},
\end{eqnarray}
where $V_{\rm osc}$ is the volume per oscillator. In each of the cases, ``strong coupling'' can be defined as the limit when the Rabi frequency $\Omega$ is much greater than $\gamma = 1/\tau$, where $\tau$ is the decay time. 

We have also seen that the standard approximation using a $2\times 2$ matrix to represent the mixing of the photon states and electronic excitation is only valid in the limit when $\omega \gg \Omega$. When $\Omega$ is comparable to $\omega$, sometimes called the ``ultrastrong'' coupling limit \cite{ultra}, the full Hamiltonian must be used. 

The use of the term ``vacuum Rabi frequency'' for the state splitting $\Omega$ is connected to the standard Rabi frequency. The standard Rabi frequency corresponds to the rate at which an electronic oscillator will flip states in the presence of a classical driving field, and is given by \cite{snokebook-rabi}
\begin{equation}
\omega_R = \frac{q|p_{cv}|}{m\hbar\omega_0} E_0,
\label{Rabi}
\end{equation}
where $E_0$ is the classical field amplitude.  To get the vacuum Rabi frequency, we use the amplitude of the field which corresponds to the average electric field amplitude in a vacuum. From (\ref{photE}) we have
\begin{equation}
\langle |E_k|^2 \rangle = \frac{\hbar\omega_0}{2\epsilon V}\langle 2a_k^{\dagger}a_k^{ }+1\rangle,
\end{equation}
which for a vacuum gives
\begin{equation}
|E_k| = \sqrt{\frac{\hbar\omega_0}{2\epsilon V}}.
\end{equation}
Then (\ref{Rabi}) becomes
\begin{eqnarray}
\omega_R &=& \frac{q|p_{cv}|}{m\hbar\omega_0} \sqrt{\frac{\hbar\omega_0}{2\epsilon V}} \nonumber\\
&=& \frac{1}{2}\sqrt{\frac{q^2}{\epsilon m}\left( \frac{2|p_{cv}|^2}{m\hbar\omega_0} \right)}\nonumber\\
&=&\frac{1}{2}\Omega,
\end{eqnarray}
which is the same as the coupling term we have used above.

One of the important physical implications of these calculation is that the coupling energy for exciton-polaritons does not depend on the number of polaritons, to lowest order. The number of excitons is not analogous to the number of atoms in the atomic case; rather, the number of excitons corresponds to the fraction of atoms in excited states. At high density, the number of excitons can affect the splitting between the upper and lower polariton branches, giving an effectively reduced coupling $\Omega$, due to phase space filling, which reduces the oscillator strength, which comes about when we can no longer use the approximation that the exciton operators are bosonic, as we did above. This has been seen \cite{2thres} to lead to a collapse of a microcavity system to weak coupling, leading to standard lasing.

\end{document}